\documentclass[twocolumn]{aastex61}

\usepackage{amsmath}
\usepackage{amssymb}

\shorttitle{Coronal Heating Topology}
\shortauthors{Rappazzo et al.}
\begin{document}

\title{Coronal Heating Topology: the Interplay of\\ Current Sheets and Magnetic Field Lines}

\author{A. F. Rappazzo}
\email{rappazzo@ucla.edu}
\affiliation{%
	Department of Earth, Planetary, and Space Sciences,
	UCLA, Los Angeles, CA 90095, USA}%
\author{W. H. Matthaeus}%
\affiliation{%
	Bartol Research Institute, Department of Physics and Astronomy, 
	University of Delaware, Newark, DE 19716, USA}%
\author{D. Ruffolo}
\affiliation{%
	Department of Physics, Faculty of Science, Mahidol University, 
	Bangkok 10400, Thailand}%
\author{M. Velli}
\affiliation{% 
	Department of Earth, Planetary, and Space Sciences,
	UCLA, Los Angeles, CA 90095, USA}%
\author{S. Servidio}
\affiliation{%
	Dipartimento di Fisica, Universit\`a della Calabria, Cosenza 
	I-87036, Italy}%



\begin{abstract}

The magnetic topology and field line random walk properties of a nanoflare-heated
and magnetically confined corona are investigated in the reduced magnetohydrodynamic regime. Field lines originating from current sheets form coherent structures, called Current Sheet Connected (CSC) regions, extended around them. CSC field line random walk is strongly anisotropic, with preferential diffusion along the current sheets' in-plane length. CSC field line random walk properties remain similar to those of the entire ensemble but exhibit enhanced mean square displacements and separations due to the stronger magnetic field intensities in CSC regions. The implications for particle acceleration and heat transport in the solar corona and wind, and for solar moss formation are discussed.

\end{abstract}




\section{Introduction} \label{sec:intro}

The stochastic properties of magnetic fluctuations in turbulent plasmas are reflected in the stochastic character of magnetic field lines, giving rise to {\it field line random walk}
\citep[FLRW,][]{1968PhRvL..21...44J, 1973ApJ...183.1029J, 1995PhRvL..75.2136M} that strongly affects the propagation and cross-field transport of energetic particles.
Additionally the intense electric fields associated to turbulent {\it coherent structures}, such as current sheets (and the related in- and out-flows), strongly contribute to particle acceleration \citep{Swann1933, 2005PhRvL..94i5001D, 2014ApJ...783..143D}.

Because in current sheets particles are energized and plasma heated, the topology of field lines that originate from them determines how these accelerated particles and heat are transported.
Furthermore in strong magnetic fields, where particle diffusion perpendicular to field lines is small and thermal conduction highly anisotropic (essentially parallel), heat and particles are in first approximation transported along field lines. 
It is therefore key to understand the interplay between current sheets and magnetic field lines.

It has become increasingly clear that the effective heating and particles acceleration occur at scales of the order of the ion (proton) inertial length $d_i$ \citep{1979anh..book...45S, 2002PhRvL..89a5002M, 2006AdSpR..38...85R, 2007PhRvL..99y5002P, 2007GeoRL..34.1101X, 2011PhPl...18k1204W, 2014GeoRL..41.4819L}, that in the solar corona, for an ion density $n_i \sim 10^8$~cm$^{-3}$, is $d_i = c/\omega_{pi} \sim 23$~m ($\omega_{pi} = \sqrt{4\pi n_i e^2/m_i}$ is the proton plasma frequency, $c$ the speed of light, $e$ the electron charge, and $m_i$ the proton mass). For typical hot coronal loops with temperatures $T \sim 10^6$~K and magnetic field intensities $B \sim 50$~G the ion gyroradius is much smaller than
$d_i$ (reaching $d_i$ only in the higher $\beta$ regions typical of the solar wind).

{\it In situ} measurements in Earth's magnetotail \citep{Runov2005,Runov2006} and magnetosheath \citep{Retino2007}, and laboratory experiments \citep{Matthaeus2005, Yamada2006} show that current sheet thickness is generally somewhat larger than the ion inertial length, with activity increasing for thinner current sheets as their width approaches $d_i$.

Additionally PIC simulations of fully developed turbulence have shown that clustering of 
current sheet thickness occurs at scales of $\sim d_i$, with substructures down to the electron inertial length $d_e$ \citep{2013PhPl...20a2303K}. These are the natural scales at which kinetic effects will convert the energy coming from large scales into the different species thermal and non-thermal energies.

Note that these findings are also consistent with the recent understanding that thin current sheets are strongly unstable under the plasmoid instability \citep{1978ZhPmR..28..193B, 1986PhFl...29.1520B, 2007PhPl...14j0703L, 2008PhRvL.100w5001L, 2009PhPl...16k2102B}, with growth rates reaching fast ``ideal'' Alfv\'en values ($\gamma \tau_A \sim 1$) for sufficiently small thicknesses \citep{2014ApJ...780L..19P, 2015ApJ...801..145T, 2015ApJ...806..131L}. Including the Hall effect the instability becomes explosive as the current sheet thickness approaches $d_i$ \citep{Pucci2017}.

Although the aforementioned studies include at most a weak guide magnetic field, the formation of current sheets with the exponentially thinning widths have been observed in fully nonlinear 2D and 3D MHD simulations
\citep{1983JCoPh..50..138S, 2002nlin......9059F, 2011PhRvE..84a6410K, 2013PhRvE..87a3110B},
and line-tied simulations with a strong guide field and vanishing initial velocity \citep{2013ApJ...773L...2R}.
Although kinetic simulations with a strong guide field are still computationally challenging,
we expect that the overall phenomenology and current sheet structure is not substantially modified in the strong guide field case of interest to the solar corona and inner heliosphere. 
{\it We then consider the gyroradii of bulk ions and electrons to be generally smaller than the current sheet thickness}, and the initial stage of {\it their acceleration is thus strongly affected by the field line topology}.

FLRW in turbulent fields is a topic of intense research \citep{Jokipii1989, Zimbardo2000, 2013ApJ...779...56S, 2013ApJ...767L..39B}, but little attention has been dedicated to the effects of spatial intermittency and coherent structures \citep{Pucci2016}.
However, the plasma thermodynamical properties are strongly affected by the topology of field lines originating in current sheets, both in the corona and solar wind. For instance the thermodynamics and high-energy radiative emission of coronal loops are determined by the temporal and spatial properties of energy dissipation along the field lines \citep{2006SoPh..234...41K, 2014LRSP...11....4R}.
Additionally, energetic particles and heat transport 
toward the transition region at coronal loops footpoints give rise to a
reticulated spongy pattern in X-rays and EUV, so-called \emph{moss}, that could
be explained by the complex trajectories of energetic particles in a stochastic
magnetic field \citep{Kittinaradorn:2009aa}.

Here we investigate the magnetic field lines random walk in a nanoflare-heated and magnetically confined corona to advance our understanding of transport of heat and energetic particles in stochastic magnetic fields, its {\it relationship with coherent structures}, and discuss its impact on coronal and solar wind dynamics.

\section{Model}

Our model coronal loops are ``straightened-out'' in a Cartesian
elongated box with axial length~L (along the z-direction)
and orthogonal square cross section of size $\ell$ (x-y planes), with aspect ratio $L/\ell$=10.
The system, with uniform density $\rho_0$,  is threaded by a strong axial magnetic field $\mathbf{B}_0 = B_0 \mathbf{\hat e}_z$, and its dynamics are well described by the reduced magnetohydrodynamic (RMHD) equations \citep{1974JETP...38..283K, 1976PhFl...19..134S, 1982PhST....2...83M, 1992JPlPh..48...85Z}, valid in the limit of a large loop aspect ratio ($\epsilon = \ell /L \ll 1$) and of a small ratio of orthogonal to axial magnetic field ($b/B_0 \le \epsilon$). The velocity ($\mathbf{u}$) and fluctuating magnetic field ($\mathbf{b}$) have only components perpendicular to the axial direction $z$, and indicating their potentials with $\varphi$ and $\psi$, they can be written as $\mathbf{u} = \nabla \varphi \times \mathbf{\hat{e}}_z$ and $\mathbf{b} = \nabla \psi \times \mathbf{\hat{e}}_z$, with the current density $j = -\nabla^2 \psi$, and vorticity $\omega = -\nabla^2 \varphi$.
In non-dimensional form they are given by:
\begin{eqnarray}
&&\partial_t \psi = \left[ \varphi, \psi \right] +B_0 \partial_z \varphi +\eta \nabla^2 \psi, \label{eq:rmhd1} \\
&&\partial_t \omega = \left[ j, \psi \right] - \left[ \omega, \varphi \right] +B_0 \partial_z j 
+\nu \nabla^2 \omega, \label{eq:rmhd2}
\end{eqnarray}
where the magnetic field has been expressed as an Alfv\'en velocity
(i.e., $b \rightarrow b/\sqrt{4\pi\rho_0}$), and then all velocities normalized to $u^\ast$ = 1~km\,s$^{-1}$ (the photospheric granulation velocity rms).
The Poisson bracket is defined as, e.g., 
$[\varphi,\psi] = \partial_x \varphi \partial_y \psi - \partial_y \varphi \partial_x \psi = - \mathbf{u} \cdot \nabla \psi$, and the Laplacian operator has only orthogonal components. Lengths are normalized to the orthogonal box length, thus $\ell$=1 and $L$=10.
Normalized resistivity and viscosity coefficients are set equal with $\eta$=$\nu$=1/$R$, where the Reynolds number 
$R$=$800$, numerical resolutions is 2048$^2\times$512, and the guide field intensity $B_0$=$10^3$. 
As in previous simulations field lines are line-tied to the top and bottom plates $z=$0,10 where they are shuffled by a prescribed  photospheric granulation-mimicking velocity constant in time with length-scale $\sim$1/4 \citep{2008ApJ...677.1348R}, while in x-y planes periodic boundary conditions are implemented.
In the x-y planes a pseudo-spectral scheme with periodic boundary conditions is implemented, time is advanced with a third-order Runge-Kutta and an adaptive time-step.
More details on the model and numerical code can be found in
\cite{2007ApJ...657L..47R, 2008ApJ...677.1348R}.

The numerical integration of Eqs.~(\ref{eq:rmhd1})-(\ref{eq:rmhd2}) cannot implement enough grid points to attain a realistic description of the internal structure of current sheets (that would additionally require the inclusion of the Hall term, or the integration of a kinetic model, that in turn could not describe properly the large scale dynamics). Nevertheless, while a good representation of the small-scale structure is very important for the acceleration of particles \citep[and we have developed a hierarchical multi-scale model to this effect in][]{2014ApJ...783..143D}, since FLRW is affected mostly by the large scale components of the magnetic field, its properties should not depend critically on the small-scale structure of the current sheets, and an MHD model represents a good starting point.
For these reasons we adopt an empirical
approach, selecting the value of resistivity for essentially
numerical reasons, but subsequently associating
the resulting dissipative scale with the ion inertial length.

\section{Results}

Our simulations start with the guide field $B_0$ (directed along~z) and no magnetic or velocity fluctuations in the computational box.
The imposed large-scale velocity at the boundaries z=0, and L twists the field lines and, once the twist exceeds a small critical threshold, the orthogonal magnetic field line tension is no longer balanced. 
Thus, as proposed by \cite{1972ApJ...174..499P, 1994ISAA....1.....P}, the magnetic field $\mathbf b$ transitions to non-equilibrium \citep{2013ApJ...773L...2R, 2015ApJ...815....8R}, bringing about turbulent dynamics that transfers energy towards the small scales where it is dissipated in \emph{nanoflares} \citep{1988ApJ...330..474P}. Line-tying keeps the velocity field in the computational box smaller than the magnetic field (far from equipartition). Nevertheless a cascade with preferential energy transfer in the x-y planes orthogonal to $B_0$ and a broad-band power-law magnetic energy spectrum develop, but the enhanced field lines stiffness introduced by line-tying gives rise to steeper magnetic energy spectra $E_M(k_\perp) \propto k_{\perp}^{-\alpha}$  with $\alpha \in [5/3, 3]$, with the steepest spectra corresponding to stronger guide fields $B_0$
\citep{1996ApJ...457L.113E, 1997ApJ...484L..83D, 1998ApJ...505..974D, 1999ApJ...527L..63D, 2003PhPl...10.3584D, 2007ApJ...657L..47R, 2008ApJ...677.1348R, 2011PhRvE..83f5401R}.
In the simulations considered here $\alpha \sim $~3.

The magnetic field structure is characterized by approximately field-aligned current sheets.
Although the overall physical conditions are markedly different between the line-tied and fully periodic reduced MHD, we hypothesize that the FLRW properties of field lines traced from current sheets are qualitatively the same, since in both cases current sheets are field aligned \citep[e.g.,][]{2005PhPl...12k2304D} and their axial extension must be linked to the parallel correlation length.
Clearly the current sheet extension in the axial direction can change depending on the particular type of forcing, but our conclusions on field line random walk and diffusion properties may be tentatively extended to the fully periodic case, as discussed in 
our Conclusions, section~\ref{sec:conclusions}.

\subsection{Field line diffusion from a single point}

\begin{figure} 
	\begin{center}
		\includegraphics[width=.9\columnwidth]{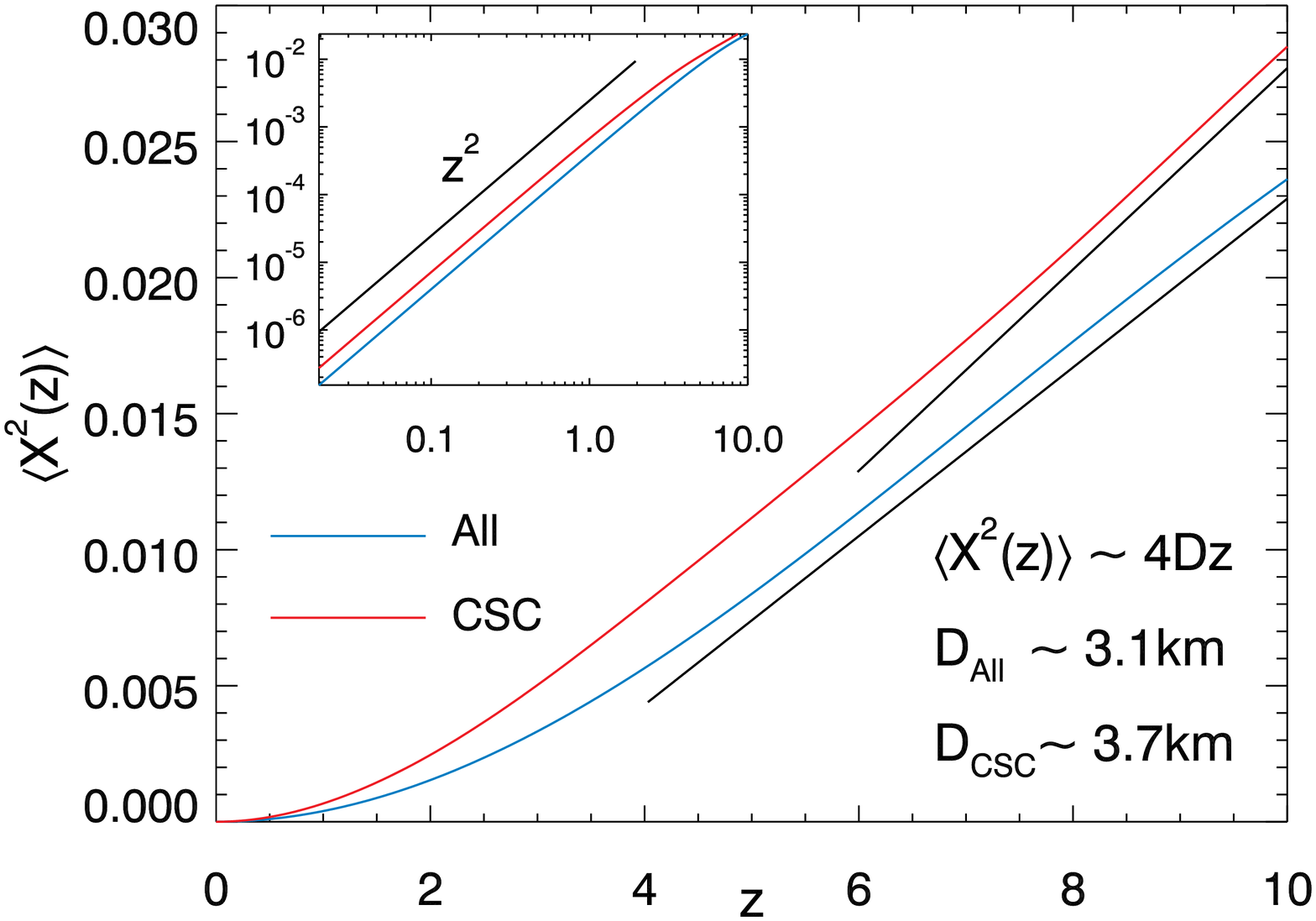}\\[1em]
		\includegraphics[width=.9\columnwidth]{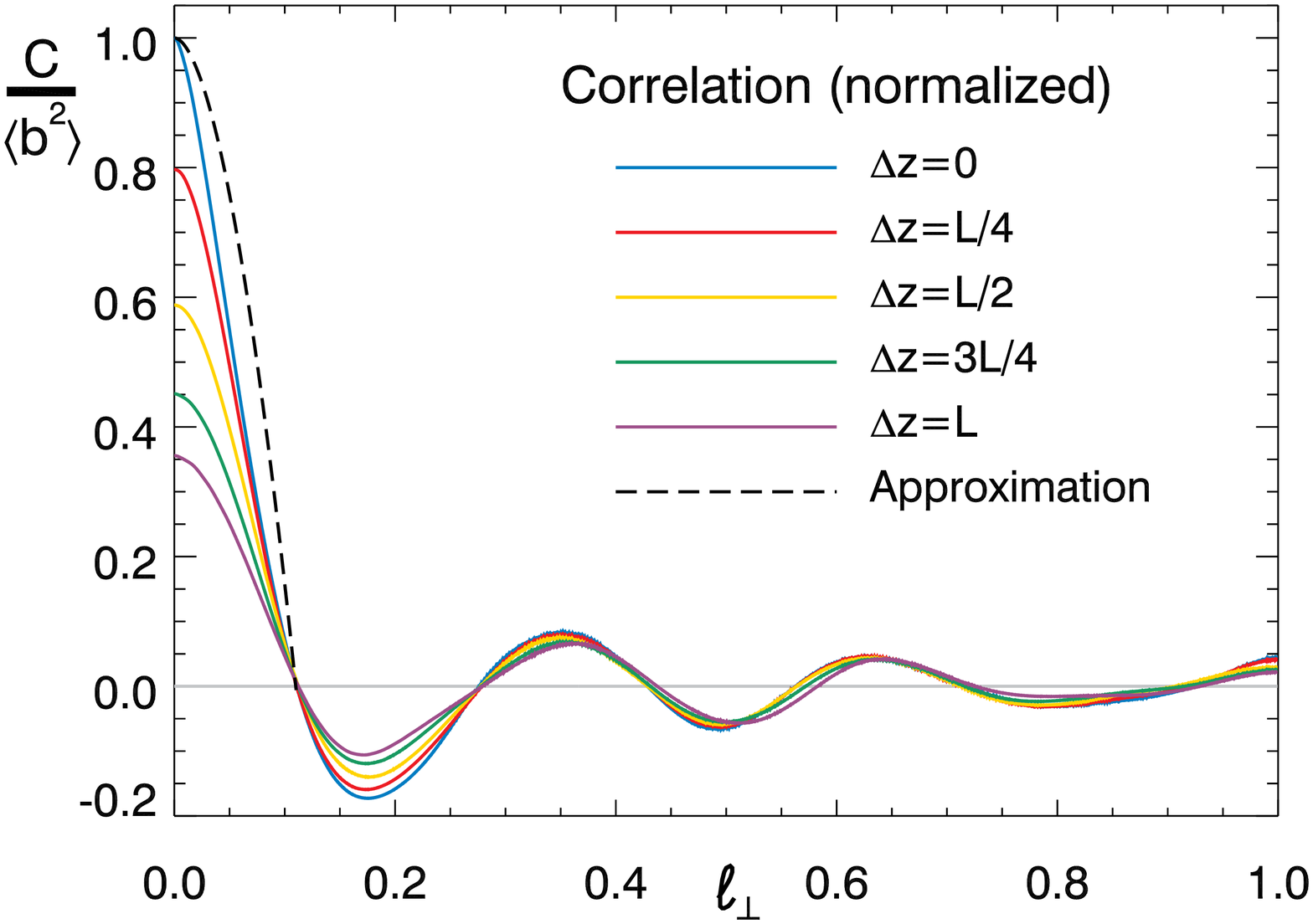}\\[1em]
		\includegraphics[width=.9\columnwidth]{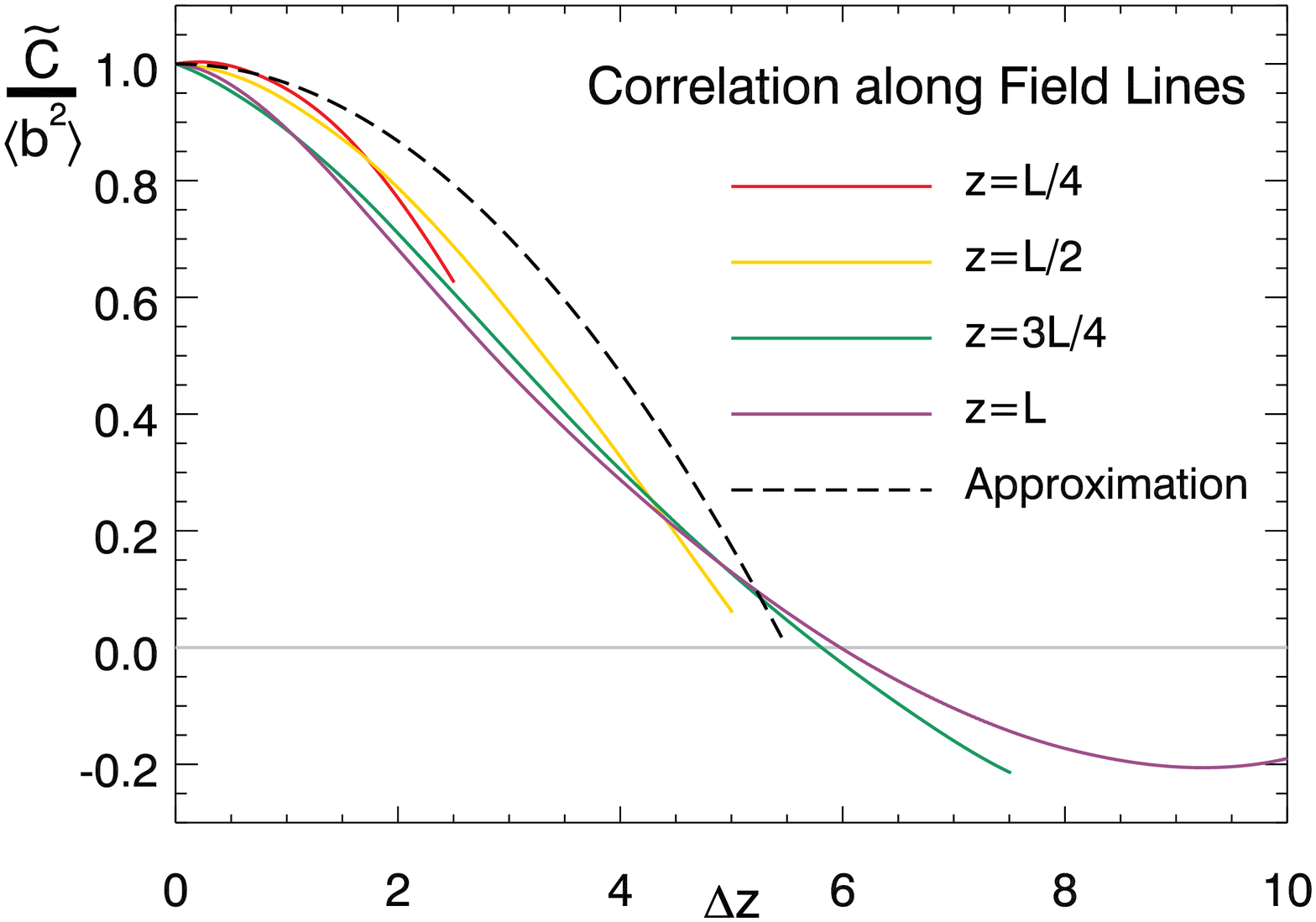}		
	\end{center}
	\caption{\label{fig:dr2}
		\emph{Top}: Mean square displacement averaged over all field lines and within the CSC region.
		\emph{Middle:} Magnetic field correlation function (Eq.~(\ref{eq:cor})) normalized with the mean square intensity $\langle b^2 \rangle$. The parallel correlation length is larger than the box size $\lambda_\parallel > L$, while $\lambda_\perp \sim 0.11$. The approximated correlation computed in Eq.~(\ref{eq:msd4}) is shown as a dashed line.
		\emph{Bottom:} Lagrangian correlation function computed along field lines (see Eq.~(\ref{eq:lcor})) as a function of $\Delta z$ for different z-values. The approximated expression from Eq.~(\ref{eq:cfl}) is shown as a dashed line.
	}
\end{figure}

Since in reduced MHD the $z$-component ($B_0$) of the magnetic
field is constant, the magnetic field line equation can be 
written as  
\begin{equation} \label{eq:lof}
\frac{\mathrm{d} \mathbf{x}}{\mathrm{d} z} \left( z \right) = 
\frac{1}{B_0}  \mathbf{b} \left( \mathbf{x}(z), z \right)
\quad \rightarrow \quad
\frac{\mathrm{d}}{\mathrm{d} z} \mathbf{x}^2 = 
\frac{2}{B_0}\, \mathbf{x} \cdot \mathbf{b},
\end{equation}
where $\mathbf{x} = (x, y)$ indicates the orthogonal coordinates.
The mean square displacement 
$\langle \mathbf{X}^2(z) \rangle = 
\langle [\mathbf{x}(z) - \mathbf{x}_0]^2 \rangle = 
\langle \mathbf{x}^2(z) \rangle - \mathbf{x}_0^2$,
where $\langle \ldots \rangle$ indicates ensemble average, 
is shown in Fig.~\ref{fig:dr2} (top panel).
After an initial ballistic stage with
$\langle \mathbf{X}^2 \rangle \propto z^2$, it subsequently
exhibits diffusion with
$\langle \mathbf{X}^2 \rangle = 4D z$.
To understand this behavior, from Eq.~(\ref{eq:lof}) we can write
\begin{equation} \label{eq:msd0}
\frac{\mathrm{d}}{\mathrm{d} z} 
\left\langle \mathbf{x}^2 (z) \right\rangle = \frac{2}{B_0^2} 
\int\limits_0^z \mathrm{d} z'\ 
\big\langle \mathbf{b} \left( \mathbf{x}(z'), z' \right) 
\cdot \mathbf{b} \left( \mathbf{x}(z),  z \right)
\big\rangle.
\end{equation}
Although the position vectors $\mathbf x (z)$ are {\it random} functions determined by the trajectory, the integrand in Eq.~(\ref{eq:msd0}) is linked to 
the magnetic field two-point \emph{correlation function}
\begin{equation}
C(\mathbf{x}_1, z_1, \mathbf{x}_2, z_2 ) =
\left\langle \mathbf{b}(\mathbf{x}_1, z_1) \cdot 
\mathbf{b}(\mathbf{x}_2, z_2) \right\rangle.
\end{equation}
For homogeneity and isotropy this depends only on the
relative parallel and orthogonal distances of the two points,
i.e., indicating with 
$\boldsymbol{\ell}_\perp = \mathbf{x}_2 - \mathbf{x}_1$,
and $\Delta z = z_2 - z_1$:
\begin{equation} \label{eq:cor}
C(\ell_\perp, |\Delta z| ) = \left\langle \mathbf{b}(\mathbf{0}, 0) \cdot 
\mathbf{b}( \boldsymbol{\ell}_\perp, \Delta z) 
\right\rangle,
\end{equation}
independent of the origination point (as long as
both points are within the z-span).
As shown in Fig.~\ref{fig:dr2} (center),
the correlation decreases at larger $\ell_\perp$ and $z$. But while it vanishes in the perpendicular direction at the correlation length $\lambda_\perp \sim 0.11$, \emph{it does not vanish in the parallel direction} (for $\ell_\perp=0$). Namely the parallel correlation length $\lambda_\parallel$ is larger than the box size $L$,
i.e., the turbulent field has a strong 2D component.
Clearly this is due to the \emph{low frequency of photospheric motions}.
Indeed, for typical hot loops, the field line footpoints are shuffled slowly 
compared to the fast Alfv\'en crossing timescale at which the induced magnetic 
field twist propagates along the loop axis.

The correlation in Eq.~(\ref{eq:msd0}) is \emph{Lagrangian}, i.e., it is computed
\emph{along the field lines}:
\begin{equation} \label{eq:lcor}
\widetilde{C} \left( z, \Delta z \right) = \langle \mathbf{b} \left( \mathbf{x}(z-\Delta z), z-\Delta z \right) \cdot \mathbf{b} \left( \mathbf{x}(z),  z \right) \rangle,
\end{equation}
with $0 \le \Delta z \le z$, and it is shown in Fig.~\ref{fig:dr2} (bottom panel).
Introducing the change of variable  $\Delta z = z - z'$ we can then write
\begin{equation} \label{eq:msd1}
\frac{\mathrm{d}}{\mathrm{d} z} 
\left\langle \mathbf{x}^2 (z) \right\rangle = \frac{2}{B_0^2} 
\int\limits_0^z \mathrm{d} \Delta z\ 
\widetilde{C}
\left( z, \Delta z \right).
\end{equation}
Since the mean square displacement between two points along a field line at a parallel distance $\Delta z$ is to a good approximation given by 
$\langle \mathbf{X}^2(\Delta z) \rangle$, the two correlations are then approximately linked by
\begin{equation} \label{eq:msd2}
\widetilde{C} \left( \Delta z \right) \sim
C( \langle \mathbf{X}^2(\Delta z) \rangle^{1/2}, \Delta z ).
\end{equation}
Additionally the correlation function is connected
to the \emph{second-order structure function} as
$C(0,0)-C(\ell_\perp,0)=
\langle \delta b_{\ell_\perp}^2\rangle/2$
\citep[e.g., see][]{2003matu.book.....B}, that in turn,
for values of $\ell_\perp$ in the
\emph{inertial range}  is linked to the
magnetic energy spectrum by
$E_{\ell_\perp} \propto \ell_\perp \delta b_{\ell_\perp}^2$, consequently
\begin{equation} \label{eq:msd3}
E_{\ell_\perp} \propto {\ell_\perp^\alpha} \quad \longrightarrow \quad 
\delta b_{\ell_\perp}^2 \propto \ell_\perp^{\alpha -1},
\end{equation}
with $\alpha \in [5/3, 3]$ for our boundary forced coronal loop model \citep{2003PhPl...10.3584D, 2007ApJ...657L..47R, 2008ApJ...677.1348R}.
Therefore extending the power-law behavior beyond the inertial range for all $\ell_\perp \le \lambda_\perp$, and taking into account that the correlation vanishes at $\lambda_\perp$, we can approximate the magnetic correlation function with
\begin{equation} \label{eq:msd4}
\frac{C(\ell_\perp,0)}{\langle b^2 \rangle} \sim 
	\begin{cases}
		1 - \left(\frac{\ell_\perp}{\lambda_\perp}\right)^{\alpha-1}
		&   \text{for}\ \ell_\perp \le \lambda_\perp,\\[1em]
		0 & \text{for}\ \ell_\perp \ge \lambda_\perp,
	\end{cases}
\end{equation}
with the exponent ranging from 2/3 for $\alpha =5/3$ up to 2 for $\alpha = 3$. This function is plotted in Fig.~\ref{fig:dr2} (middle) for $\alpha = 3$. Here we neglect the parallel variation of $C$ when used in Eq.~(\ref{eq:msd2}) because $\langle \mathbf{X}^2(\Delta z) \rangle$ increases monotonically with $\Delta z$, and as shown in Fig.~\ref{fig:dr2} (middle) the curves then tend to overlap quickly becoming approximately independent of $\Delta z$.

\begin{figure} 
	\centering
	\includegraphics[width=.95\columnwidth]{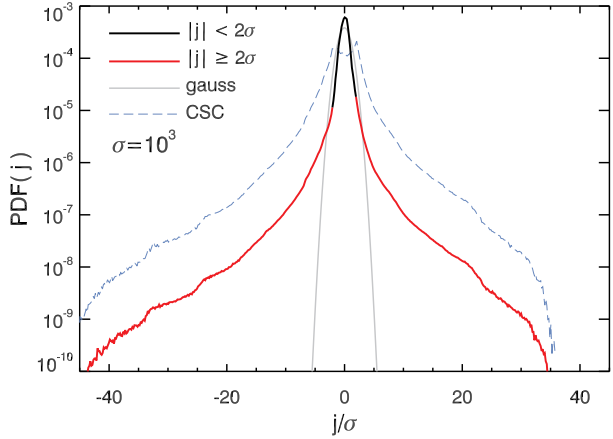}		
	\caption{\label{fig:pdf}
		Probability density function (pdf) of current density. Current sheets are defined as the spatial regions where $|j| \ge 2\sigma$, with $\sigma = \langle j^2 \rangle^{1/2}$.
		For reference we plot also the gaussian distribution with same standard deviation $\sigma$, and the pdf of $j$ computed in the CSC region (dashed line).
	}
\end{figure}
\begin{figure*}
	\begin{center}
		\includegraphics[width=.32\textwidth]{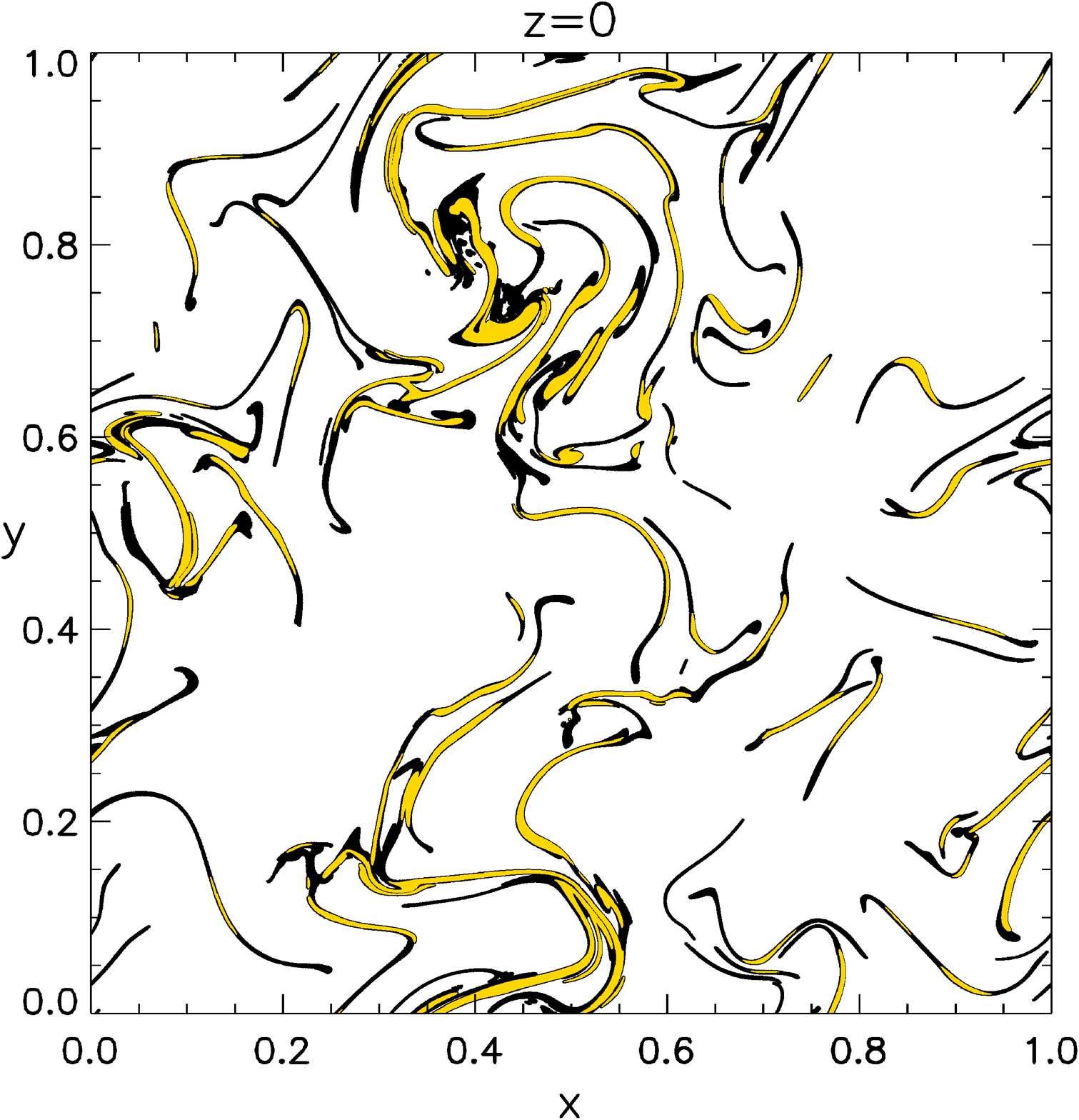}
		\hspace{0.008\textwidth}
		\includegraphics[width=.32\textwidth]{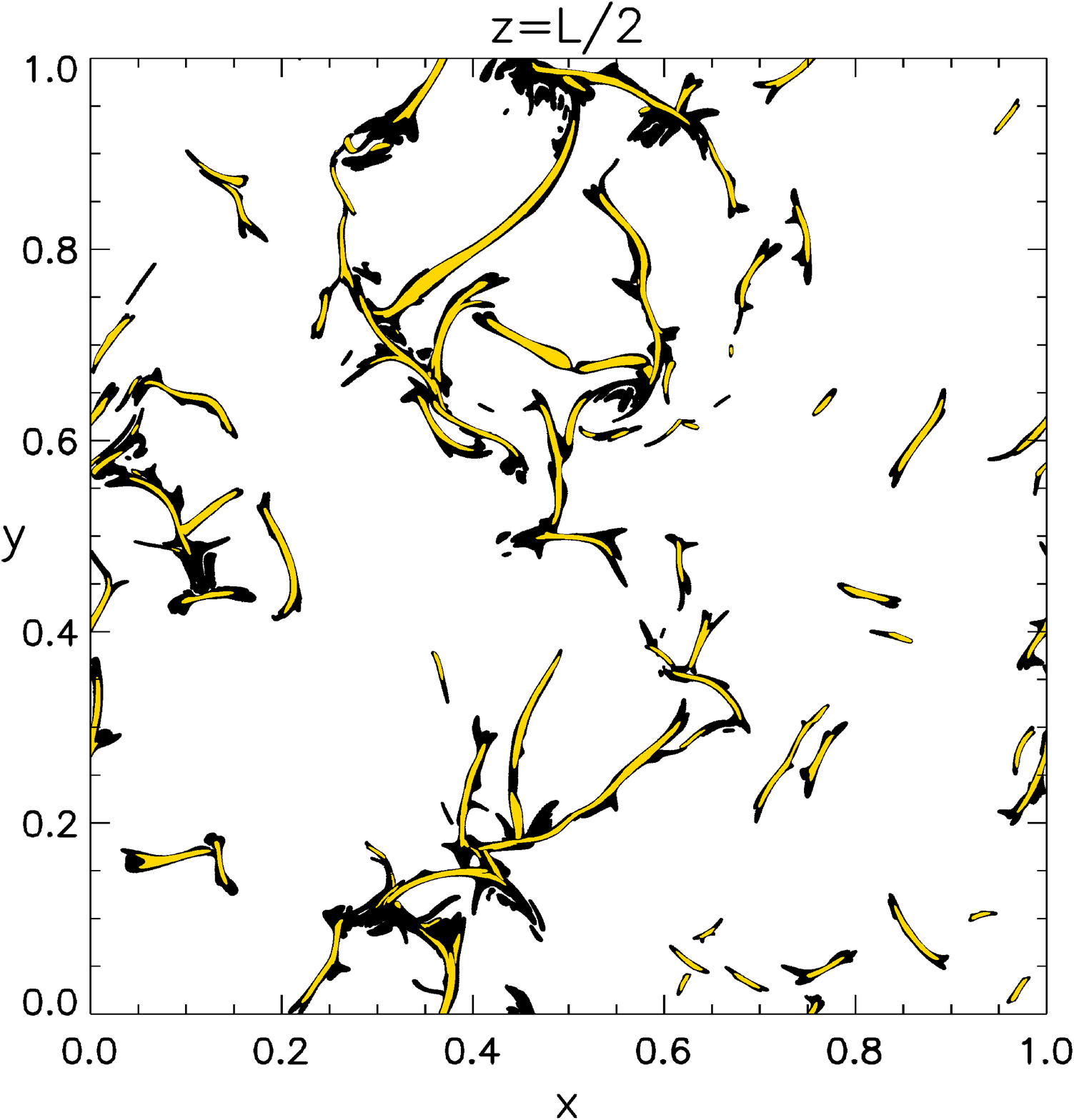}
		\hspace{0.01\textwidth}
		\includegraphics[width=.32\textwidth]{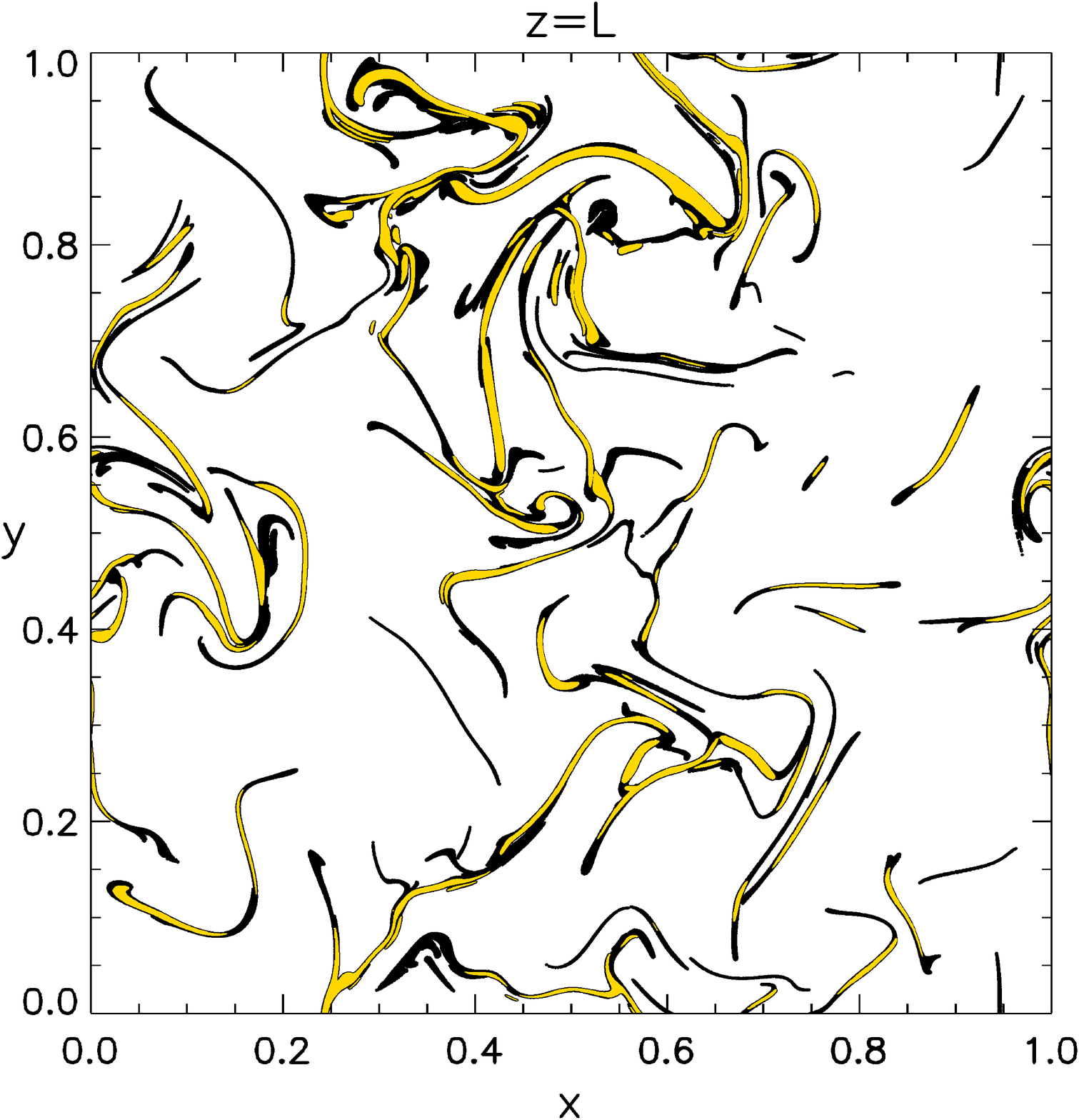}
	\end{center}\vspace{-1em}
	\caption{\label{fig:csl}
		Current sheets (defined as those locations where $|j| \ge 2\langle j^2 \rangle^{1/2}$) are shown in yellow
		in the bottom, middle and top plates (z=0, L/2, and L, with axial length L=10).
		The locations where the field lines traced from current sheets in other planes cross the present plane outside the current sheet are drawn as black dots. Together both regions form the Current Sheet Connected (CSC) region. {\it This figure, showing additional planes, is available online as an animation.}}
\end{figure*}

The behavior of the mean square displacement $\langle \mathbf{X}^2(z) \rangle$
can then be readily understood from the correlation function
$C(\ell_\perp,\Delta z)$ (Fig.~\ref{fig:dr2}).
For small values of $z$ the integral in Eq.~(\ref{eq:msd1})
can be Taylor-expanded, and since 
$C(0,0) = \langle b^2 \rangle$, to the first order we obtain
$\langle \mathbf{X}^2(z) \rangle \approx 
z^2 \langle b^2 \rangle/B_0^2$.
On the other hand as soon as 
$\langle \mathbf{X}^2(z) \rangle^{1/2} \gtrsim \lambda_\perp$ 
exceeds the orthogonal correlation length 
$\lambda_\perp \sim 0.11$,
the integral in Eq.~(\ref{eq:msd1}) remains approximately
constant, because the largest contribution comes from 
$\ell_\perp < \lambda_\perp$, hence 
$\langle \mathbf{X}^2 (z) \rangle \sim 4Dz$ diffuses linearly.
The transition from the ballistic to the diffusive stage occurs for
$\langle \mathbf{X}^2(z) \rangle \sim z^2 \langle b^2 \rangle / B_0^2 \sim \lambda_\perp^2 $, i.e., for $z_D \sim \lambda_\perp B_0/\langle b^2 \rangle^{1/2}$. In our case, since $\lambda_\perp \sim 0.11$, $B_0=10^3$, and $\langle b^2 \rangle^{1/2} \sim 20$, the transition occurs at $z_D \approx 5.5$, as confirmed in Fig.~\ref{fig:dr2} (top).

We can estimate the diffusion coefficient $D$ by using in Eq.~(\ref{eq:msd2}) the approximation for the mean square displacement outlined in the previous paragraph (i.e., $\langle \mathbf{X}^2(\Delta z) \rangle \approx \Delta z^2 \langle b^2 \rangle /B_0^2$ for $z \le z_D$). The Lagrangian correlation along the field lines can then be approximated from Eqs.~(\ref{eq:msd2}) and (\ref{eq:msd4}) with
\begin{equation} \label{eq:cfl}
\frac{\widetilde{C}(\Delta z)}{\langle b^2 \rangle} =
	\begin{cases} 
		1- \left( \frac{\Delta z}{z_D} \right)^{\alpha -1}
		& \text{for }\  \Delta z \le z_D,\\[1em]
		0 & \text{for }\  \Delta z \ge z_D.
	\end{cases}
\end{equation}
Substituting in Eq.~(\ref{eq:msd1}), and integrating it to 
obtain $\langle \mathbf{X}^2(z) \rangle$, the diffusion 
coefficient $D$ is then given by
\begin{equation} \label{eq:diff}
D \sim  
\frac{\alpha -1}{2\alpha}\ 
\lambda_\perp
\frac{\langle b^2\rangle^{1/2}}{B_0},
\end{equation}
a functional form characteristic of Bohm diffusion \citep{2011ApJ...741...16G}, with the coefficient $(\alpha-1)/2\alpha$ ranging in the narrow interval $[1/5, 1/3]$ as $\alpha \in [5/3, 3]$.
Since $\langle b^2 \rangle^{1/2} \sim 20$, $B_0=10^3$, $\lambda_\perp \sim 0.11$
and $\alpha \sim 3$, we obtain $D \sim 7.3\times 10^{-4}$, corresponding to 
$D \sim 2.9$\,km in conventional units, compatible with $D \sim 3.1$\,km
computed from our simulation (Fig.~\ref{fig:dr2}, top panel).
The approximated Lagrangian correlation function $\widetilde{C} (\Delta z)$ (Eq.~(\ref{eq:cfl})) with $\alpha=3$ and $z_D \sim 5.5$ is shown in Fig.~\ref{fig:dr2} (bottom panel).

\subsection{Current sheet connected regions}

The current density $j$ is intermittently distributed in space, as typical of turbulent systems. Its Probability density function (pdf), shown in Fig.~\ref{fig:pdf}, is not gaussian and exhibits typical large tails where the current is strong, corresponding to current sheets in physical space. The noticeable skewness in Fig.~\ref{fig:pdf} results from the use of a single snapshot, and it is a fluctuation that vanishes when averaging over several snapshots, i.e., the time-averaged distribution is symmetric. 

We define as current sheets all those spatial regions where current is larger than two standard deviations $|j| \ge 2\sigma$, with $\sigma = \langle j^2 \rangle^{1/2}$,
shown in yellow for the representative planes z=0, L/2 and L in Fig.~\ref{fig:csl}.

The relationship between magnetic field topology and current sheets is then investigated by tracing field lines originating in current sheets.
Specifically field lines are traced from all the grid points where $|j| \ge 2\sigma$ in 9 equispaced x-y planes (from z=0 up to z=L, separated by L/8). For grid points that are not at the boundaries z=0 or z=L the respective field lines are traced both forward and backward with respect to the z-direction.
We trace a total of 1,107,242 field lines, extending from the bottom to the top plate, and their intersection with the selected plane is shown as a black dot in Fig.~\ref{fig:csl}.
Clearly field lines are present also in the yellow regions, both those traced from there plus others originating from current sheets in different planes.

Current sheets are elongated in the guide field direction $z$, and the field lines traced from them form similarly shaped \emph{coherent structures}, that we indicate as \emph{Current Sheet Connected} (CSC) \emph{regions}. Although current sheets in reduced MHD have a complex structure with a cross-shear magnetic field component and mostly external X-points \citep{2013ApJ...771..124Z, 2014ApJ...797...63W},
noticeably the presence of a strong magnetic shear in correspondence of current sheets makes the field line random walk \emph{strongly anisotropic}, with field lines diffusing preferably along the in-plane sheet length and very little across it (Fig.~\ref{fig:csl}).
Since diffusion increases with distance its effects are most apparent in planes z=0 and L.

The coherence and strong anisotropy of the CSC regions are in stark contrast with the homogeneity of the stochastic properties typically associated with FLRW, and their well-know tendency to fragment flux tubes \citep{1995PhRvL..75.2136M,2014ApJ...785...56S,2012ApJ...758L..14R}. 
Nevertheless the mean square displacement $\langle \mathbf{X}^2(z) \rangle$ of the CSC field lines has properties similar to those of the entire ensemble, as shown in Fig.~\ref{fig:dr2} (top). The higher value of the diffusion coefficient  $D_{\rm CSC}/D_{\rm All} \sim 1.2$ is due to the higher magnetic field intensity in the CSC region as $\langle b^2 \rangle_{CSC}^{1/2} / \langle b^2 \rangle_{All}^{1/2} \sim 1.2$ in agreement with Eq.~(\ref{eq:diff}).

\subsection{Pair separation}

\begin{figure*}
	\includegraphics[width=1\columnwidth]{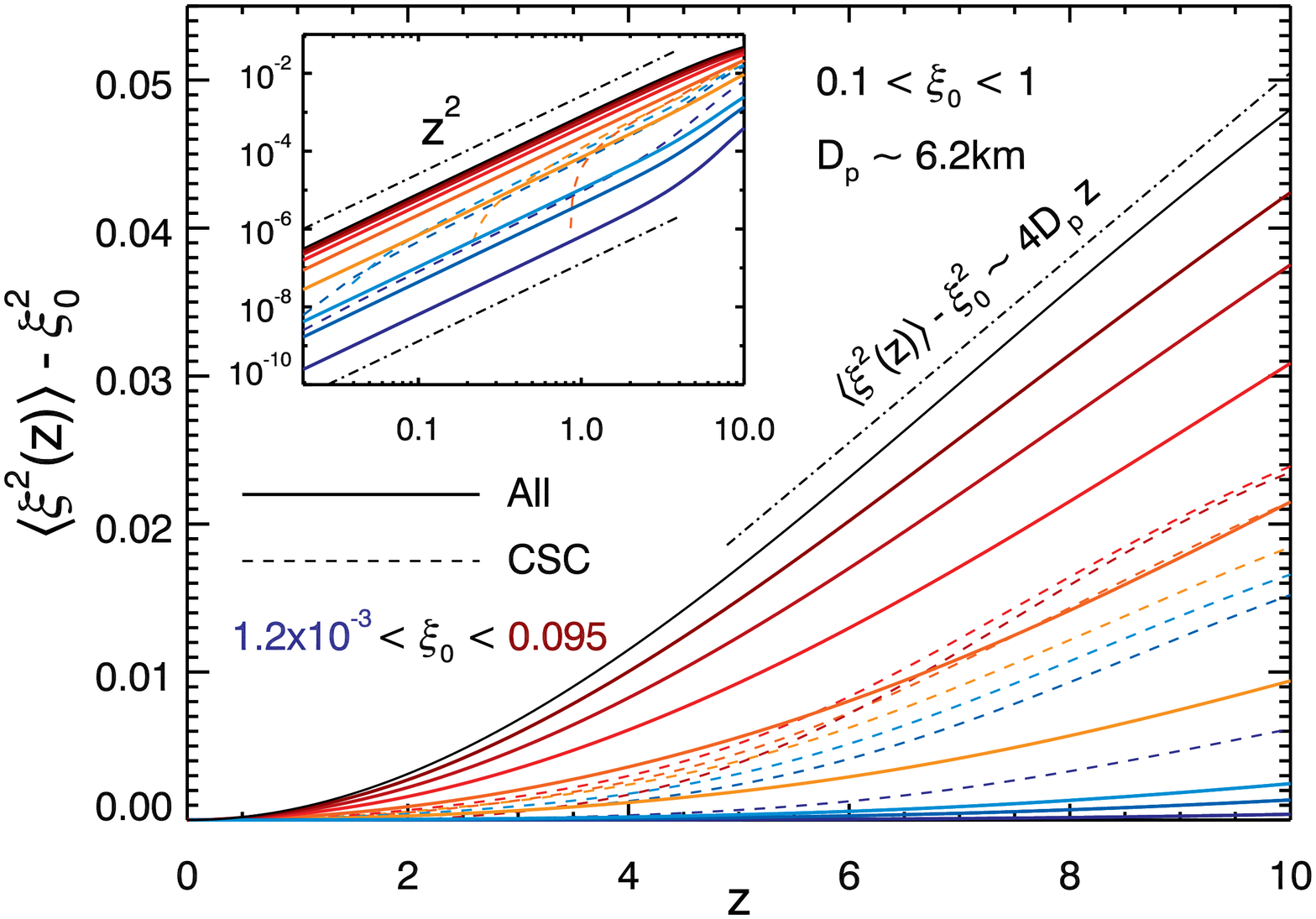}\hspace{1em}
	\includegraphics[width=1\columnwidth]{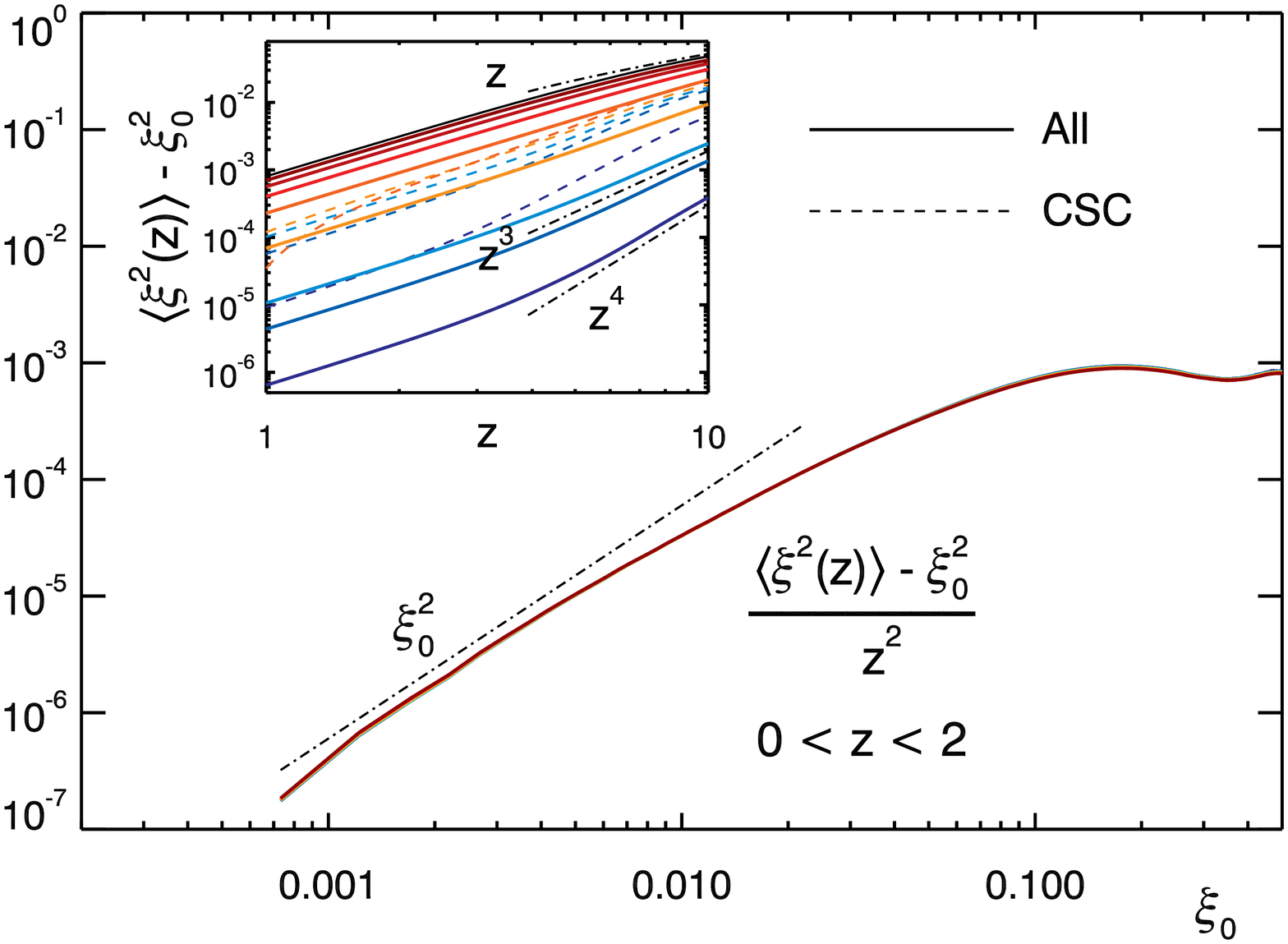}
	\caption{\label{fig:psav} \emph{Left:} Mean square separation of field line pairs is plotted against axial distance $z$ for different initial separations $\xi_0$ (\emph{continuous lines} consider the whole field lines ensemble). Each red continuous line represents a bin of initial separation 0.01 wide centered at $\xi_0=0.015, 0.035, .., 0.095$. Since all curves with
	$0.1 < \xi_0 < 1$ overlap, we draw their average with a black continuous line. The 3 blue continuous lines represents bins with grid resolution width 
	dx=1/2048 centered at $\xi_0= 2.5, 6.5$, and $10.5$\,dx corresponding to $\xi_0 = 1.2, 3.2$, and $5.2 \times 10^{-3}$.
	\emph{Insets} in both panels show logarithmic plots of mean square separations, the right inset is a magnification of the left one showing the development of superdiffusion.
	\emph{Dashed lines} show CSC field line separations with red and blue lines centered at same bins as the corresponding curves for the all ensemble, and with bin width dx. The red line at $\xi_0=0.095$ is not drawn for the CSC case due to low statistics.
	\emph{Right:} $(\langle \xi^2  (z) \rangle - \xi^2_0)/z^2$ is plotted against $\xi_0$ for different values of $z \in [0,2]$ showing that it scales approximately as $\xi_0^2$ in the inertial range.}
\end{figure*}

To further understand the magnetic topology we consider 
the separation of field line pairs. From Eq.~(\ref{eq:lof})
their orthogonal separation in the $x$--$y$ plane 
$\boldsymbol{\xi} (z) = \mathbf{x}_2(z) - \mathbf{x}_1(z) $,
given the initial separation $\boldsymbol{\xi}(0) = \boldsymbol{\xi}_0$,
is determined as a function of $z$ by
\begin{equation} \label{eq:los}
\frac{\mathrm{d} \boldsymbol{\xi}}{\mathrm{d} z}(z) = 
\frac{1}{B_0} 
\Big[ \mathbf{b} \left( \mathbf{x}_2(z), z \right)
-\mathbf{b} \left( \mathbf{x}_1(z), z \right) \Big].
\end{equation}
Similarly to the single field line case we obtain
\begin{eqnarray} \label{eq:los2}
\frac{\mathrm{d} \langle \boldsymbol{\xi}^2(z) \rangle
}{\mathrm{d} z} = \frac{2}{B_0^2}
\int\limits_0^z \mathrm{d} z' 
&\Big\langle \big[& \mathbf{b} \left( \mathbf{x}_2(z'), z' 
\right)
-\mathbf{b} \left( \mathbf{x}_1(z'), z' \right) \big]\qquad \\
\cdot &\big[& \mathbf{b} \left( \mathbf{x}_2(z), z \right)
-\mathbf{b} \left( \mathbf{x}_1(z), z \right) \big] 
\Big\rangle,\nonumber
\end{eqnarray}
that following \cite{2004ApJ...614..420R} can be written as
\begin{eqnarray}
&&\frac{\mathrm{d}}{\mathrm{d} z}  \label{eq:los3} 
\langle \boldsymbol{\xi}^2(z) \rangle = \frac{4}{B_0^2}
\left( I_{11} - I_{12}  \right), \quad \mathrm{where:}\\
&&I_{11} = \int\limits_0^z \mathrm{d} z'\ \label{eq:los31}
\big\langle \mathbf{b} \left( \mathbf{x}_1(z'), z' \right)
\cdot \mathbf{b}\left(\mathbf{x}_1(z),z\right)\big\rangle,\\
&&I_{12} = \int\limits_0^z \mathrm{d} z'\ \label{eq:los32}
\big\langle \mathbf{b} \left( \mathbf{x}_1(z'), z' \right)
\cdot \mathbf{b} \left( \mathbf{x}_2(z), z \right) \big\rangle.
\end{eqnarray}
$I_{11}$ is same as the integral in Eq.~(\ref{eq:msd0}) because
it refers to single field lines, therefore \emph{we already
	understand its behavior}.
But $I_{12}$ differs as it considers a pair.

As indicated by \cite{1950QJRMS..76..133B} for the
hydrodynamic case, for $z$ sufficiently small the mean-square 
separation will grow quadratically with z 
(see inset in Fig.~\ref{fig:psav}, left panel) and it will be 
proportional to the \emph{second-order structure function}.
Indeed from Eq.~(\ref{eq:los3}), Taylor-expanding Eqs.~(\ref{eq:los31})-(\ref{eq:los32}) in z, and since 
$C(0,0)-C(\mathbf{\xi}_0,0)= \langle \delta b_{\xi_0}^2\rangle/2$
we obtain 
\begin{eqnarray}
&&\frac{\mathrm{d}}{\mathrm{d} z} \langle \boldsymbol{\xi}^2 (z) \rangle =
2z\ \frac{\langle \delta b_{\xi_0}^2 \rangle}{B_0^2},
\quad \mathrm{hence}\\
&&\langle \boldsymbol{\xi}^2 (z) \rangle - \boldsymbol{\xi}^2_0 =
z^2\ \frac{\langle \delta b_{\xi_0}^2 \rangle}{B_0^2},
\end{eqnarray}
where as usual $\xi_0 = |\boldsymbol{\xi}_0|$. 
Additionally we can now approximate the second order structure function as in Eq.~(\ref{eq:msd4}) with 
$\langle \delta b^2_{\xi_0} \rangle \sim 2 \langle b^2 \rangle (\xi_0/\lambda_\perp)^{\alpha-1}$ for $\xi_0 \le \lambda_\perp$ and 
$\langle \delta b^2_{\xi_0} \rangle \sim 2 \langle b^2 \rangle$ for
$\xi_0 \ge \lambda_\perp$, with the scaling relation more accurate
for values of $\xi_0$ in the \emph{inertial range}.
We can then write
\begin{equation} \label{eq:los4}
\langle \boldsymbol{\xi}^2 (z) \rangle - \boldsymbol{\xi}^2_0 \approx
	\begin{cases}
		\frac{2}{\lambda_\perp^{\alpha-1}}\, \frac{\langle b^2 \rangle}{B_0^2}\ 
		z^2\, \xi_0^{\alpha-1}, & \quad \mathrm{for}\ \xi_0 \le \lambda_\perp\\[1em]
		2 \frac{\langle b^2 \rangle}{B_0^2}\ z^2, & \quad \mathrm{for}\ \xi_0 \ge \lambda_\perp\\		
	\end{cases}
\end{equation}
with $\alpha \in [5/3,3]$.

We plot $(\langle \boldsymbol{\xi}^2  (z) \rangle - \boldsymbol{\xi}^2_0)/z^2$  in Fig.~\ref{fig:psav} (right panel) as a function of $\xi_0$ for 11 values of $z \in [0,2]$ separated by 0.2, almost perfectly overlapping and showing that in the inertial range ($\xi_0 \lesssim 0.1$) it scales approximately as $\xi_0^{2}$, compatible with $\alpha \sim 3$ in our simulations, and saturates correctly to $\sim 2\langle b^2 \rangle/B_0^2 \sim 8 \times 10^{-4}$ with our parameters ($\langle b^2 \rangle \sim 400$, $B_0=10^3$, $\lambda_\perp \sim 0.11$). 
The small departure from $\xi_0^{2}$ in the inertial range shown in Fig~\ref{fig:psav}
occurs because the second order structure function 
$\langle \delta b_{\xi_0}^2/2 \rangle = C(0,0)-C(\xi_0,0)$ considers zero separation in z and is calculated in the bottom boundary plane $z=0$ (since the coefficients in the Taylor expansion in z of Eqs.~(\ref{eq:los31})-(\ref{eq:los32}) are calculated for z=0). Indeed the 2D magnetic energy spectrum averaged over the whole box $E_{\ell_\perp} \propto \ell_\perp \delta b_{\ell_\perp} \sim \ell_{\perp}^3$, corresponding to $\delta b_{\ell_\perp} \sim \ell_{\perp}^2$, and the same behavior is observed for the spectra in all z-planes, except those in proximity of the boundaries z=0 and L where line-tying boundary conditions are applied. At these boundaries the velocity field is prescribed and therefore the dynamics does not follow the same equations as in the interior, with the effect of slightly modifying the magnetic energy spectrum for the planes in their close proximity.
Nevertheless \emph{the scaling} for the separation remains very close to a $\propto \xi_0^2$ scaling, \emph{departing strongly in the inertial range from $\xi_0^{2/3}$ expected for a standard Kolmogorov spectrum with $\alpha = 5/3$} \citep{1950QJRMS..76..133B,2006Sci...311..835B}. Therefore the ratio of mean square separations for field lines with relative larger initial separations $\xi'_0 > \xi_0$ are increasingly bigger for  steeper spectral indices, indeed from Eq.~(\ref{eq:los4}) the ratio of their separations in the ballistic range ($\propto z^2$) grows like $(\xi'_0/\xi_0)^{\alpha-1}$.

In general, as $\xi_0 \rightarrow 0$ the two field lines 
tend to the same field line, i.e., 
$\mathbf{x}_2(z) \rightarrow \mathbf{x}_1(z)$,
consequently $I_{12} \rightarrow I_{11}$,
and the mean square separation vanishes in this limit.
On the other hand the mean square separation in $I_{12}$, i.e.\
$\xi_{12}(z',z)=\langle [\mathbf{x}_2(z') - \mathbf{x}_1(z)]^2 \rangle^{1/2}$
is always larger than $\xi_0$. For initial separations larger than $\lambda_\perp$
the correlation is small $\forall z \in [0,L]$  so that $I_{12} \approx 0$.
In this case the diffusion coefficient for 
pair separation is double that of single field
line diffusion, i.e., 
$\langle \boldsymbol{\xi}^2 (z) \rangle - \boldsymbol{\xi}^2_0 = 4D_{\mathrm{p}}z$
with $D_{\mathrm{p}} = 2D$, as shown in Fig.~\ref{fig:psav} (left panel) for $0.1 < \xi_0 < 1$ by the continuous black line (we average these curves since they overlap).
For any $\xi_0 < \lambda_\perp$ there is always a critical height
$z_D(\xi_0)$ above which the mean separation between the two field 
lines in $I_{12}$ is larger than $\lambda_\perp$.
Hence the increasingly larger negative contribution of the $I_{12}$ term 
to mean square pair separation will display diffusion at progressively lower heights,   with a smaller total diffusion coefficients $D \le D_p \le 2D$, 
as shown in Fig.~\ref{fig:psav} by the red lines, that consider 5 bins
with $\xi_0 \in [0.015,0.095]$ with $\Delta \xi_0 =0.02$.
Clearly, for sufficiently small initial separations, field lines will not be able to display diffusion because our system is bounded in the axial direction~z to maximum length L and separation cannot grow up to the perpendicular correlation length, as shown well by the blue continuous lines in Fig.~\ref{fig:psav}. 

A ballistic stage $\propto z^2$ is always present initially for $z \lesssim 3$, as shown in Fig.~\ref{fig:psav} (insets), but field lines with initial separation within the \emph{dissipative range} ($\xi_0 \lesssim 10$\,dx, where dx=1/nx=1/2048 is the numerical grid step size) exhibit subsequently a Richardson-like superdiffusive stage \citep{2013ApJ...767L..39B,2013Natur.497..466E,Servidio2016}
with mean square separation up to $\propto z^4$ (blue continuous lines in Fig.~\ref{fig:psav}, with respectively $\xi_0=2.5,$ 6.5, and 10.5\,dx), while for larger separations they transition to the diffusive regime.

Even though FLRW in CSC regions is strongly anisotropic, CSC field line pair separation (shown in Fig.~\ref{fig:psav} with dashed lines) exhibit similar properties to the whole ensemble (shown with continuous lines, color code is the same for both line types).
Their statistics are degraded for larger $\xi_0$ because in CSC regions the number of field line couples diminishes at larger separations so that the averages do not saturate yet to their ensemble value. This can be seen in the left inset, where the red dashed curves become negative, because the relation 
$\langle \boldsymbol{\xi}^2(z) \rangle - \boldsymbol{\xi}_0^2 = \langle (\boldsymbol{\xi}(z)-\boldsymbol{\xi}_0)^2 \rangle \ge 0$ is valid only for a sufficiently high number of field lines, when
$\langle \boldsymbol{\xi}(z) \rangle = \boldsymbol{\xi}_0$.

The main difference between CSC pair separations and the whole ensemble is that for same
initial separation $\xi_0$ the CSC field lines exhibit higher separations. Similarly to mean square displacement (Fig.~\ref{fig:dr2}, top panel) the higher values are due to the greater magnetic field intensity in CSC regions 
($\langle b^2 \rangle_{CSC}/\langle b^2 \rangle \sim 1.5$), and indeed the separation in the ballistic range is proportional to $\langle b^2 \rangle$ (Eq.~(\ref{eq:los4})).
Notice that the coefficient in Eq.~(\ref{eq:los4}) includes also the perpendicular correlation length $\lambda_\perp$ that is not readily computable in the non-Cartesian CSC region, but from the data we can estimate that the separation for field lines with same initial separation is about 10 times larger in CSC regions than for the whole ensemble. 
As mentioned previously the statistics are degraded for CSC field lines with larger initial separations. Therefore while the larger separations for CSC field lines is very well demonstrated for smaller initial separations (blue dashed lines, and first red dashed line), we cannot yet fully verify this conclusion for larger initial separations, as the averages have not yet saturated to their ensemble values (this point will be further investigated in upcoming work). 

\section{Conclusions and Discussion} \label{sec:conclusions}

To gain insight into particle acceleration and heat transport in coronal loops, solar wind, and more in general for plasmas in the reduced MHD regime, we have investigated the magnetic topology of field lines originating from current sheets.
We have found that they form coherent structures, dubbed Current Sheet Connected (CSC) regions, similarly to the current sheets they originate from.
Field lines in these regions perform highly anisotropic FLRW, with diffusion occurring preferentially along the current sheet in-plane length. Nevertheless FLRW and diffusion coefficients have similar properties for CSC field lines and the whole ensemble, with larger displacement and separations occurring in CSC regions where the magnetic field intensity is higher.

This emerging picture has strong implications for particle acceleration and heat transport, particularly in the low corona, where all protons and electrons with temperatures below $10^6$\,K have gyroradii smaller than the current sheet thickness.
It indeed implies that in coronal loops particle and heat are transported almost exclusively within the CSC region, a small volume of plasma around current sheets with a small filling factor, while most of the volume is topologically disconnected from current sheets and the associated flow of particles and heat.

This picture is fully consistent with observations \citep{2010ApJ...723.1180S, 2008ApJ...686L.131W, 2009ApJ...695..642U, 2009ASPC..415..221K} and recent thermodynamical 3D simulations \citep{2012A&A...544L..20D, 2016ApJ...817...47D}
strongly suggesting that coronal loops cannot be modeled with single isothermal flux tubes, as their radiative properties can only be explained by the presence of both hot and cold plasmas at observational sub-resolution scales (multi-temperature loops).
Also, the complex topology in CSC regions, with enhanced magnetic field line displacements and separations, points to a complex stochastic nature for the heating function along the field lines.

Additionally the structure of the CSC regions in the top and bottom plates z=0 and L
is consistent with that of so-called \emph{moss}, the spongy reticulated pattern in X-rays and EUV formed at the coronal base of hot loops, confirming that the FLRW can play a strong role in the formation of these structures as recently proposed \citep{Kittinaradorn:2009aa}.

The reduced MHD FLRW properties and topology strongly support the results of recent test-particle simulations \citep{2014ApJ...783..143D} with initial gyroradii smaller than current sheet widths  propagated in similar magnetic fields to those discussed here.
Particles are at first strongly accelerated along the z-direction in current sheets by the strong electric field associated with the current, until they pitch-angle scatter thus increasing their gyroradii. Subsequently, as long as the gyroradius is smaller than the orthogonal correlation length, those that remain close to the CSC region are accelerated by a (non-magnetic moment conserving) betatron-like mechanism due to the inhomogeneous $\mathbf{u}\times B_0 \mathbf{\hat{z}}$ electric field associated to outflows in current sheets.

For parameters typical of hot solar coronal loops we have found that the parallel correlation length is longer than the loop length, and our results consider this specific case. Nevertheless they may also apply to the unbounded (e.g., periodic) case, which we plan to investigate thoroughly in upcoming work. In general we expect the current sheet length along z to be strongly correlated with the magnetic field parallel correlation length $\lambda_\parallel$.
Therefore we expect that CSC regions connected to any such current sheet of length $\sim \lambda_\parallel$ to have a similar structure to those found here around the current sheet. But as field lines are traced further away at distances larger than $\lambda_\parallel$ we expect the CSC region to fragment and the associated FLRW to lose anisotropy and acquire more homogeneous properties, i.e., the CSC field lines will at that point connect and diffuse isotropically throughout the volume (hence mostly in regions with low current).

Therefore in unbounded systems we expect heat and accelerated particles to be initially confined to CSC regions around current sheets (of length $\sim \lambda_\parallel$), but further away heat and particles would distribute more uniformly throughout the plasma.
This picture is strongly consistent with recent analyses of solar wind data \citep{2012PhRvL.108z1102O}, where temperature is found to peak in regions with high magnetic field gradients, while it rapidly descends to approach the ambient solar wind temperature as distance from those regions increases.

\acknowledgments
Resources supporting this work were provided by the NASA High-End Computing (HEC) Program through the NASA Advanced Supercomputing (NAS) Division at Ames Research Center. This work was partially supported by NASA through Grand Challenge Research grant NNX15AB88G, Living with a Star grant NNX14AI63G, as well as the Solar Probe Plus Observatory Scientist grant, and by the Thailand Research Fund (grant RTA5980003).

\end{document}